\documentclass[10pt,letterpaper]{article}

\usepackage{cogsci}

\cogscifinalcopy 

\usepackage{graphicx}
\usepackage{amsfonts}
\usepackage{pifont}
\usepackage{multirow}

\usepackage{xcolor}

\usepackage{pslatex}
\usepackage{apacite}

\title{The Wisdom of Intellectually Humble Networks}
 
\author{{\large \bf Mohammad Ratul Mahjabin (mohammadratul@usf.edu)} \\
  Bellini College of Artificial Intelligence, Cybersecurity and Computing, University of South Florida \\
  4202 E Fowler Avenue, Tampa, FL-33620, USA
  \AND {\large \bf Raiyan Abdul Baten (rbaten@usf.edu)} \\
  Bellini College of Artificial Intelligence, Cybersecurity and Computing, University of South Florida \\
  4202 E Fowler Avenue, Tampa, FL-33620, USA}

\begin{document}

\maketitle

\begin{abstract}
People's collectively shared beliefs can have significant social implications, including on democratic processes and policies. Unfortunately, as people interact with peers to form and update their beliefs, various cognitive and social biases can hinder their collective wisdom. In this paper, we probe whether and how the psychological construct of intellectual humility can modulate collective wisdom in a networked interaction setting. Through agent-based modeling and data-calibrated simulations, we provide a proof of concept demonstrating that intellectual humility can foster more accurate estimations while mitigating polarization in social networks. We investigate the mechanisms behind the performance improvements and confirm robustness across task settings and network structures. Our work can guide intervention designs to capitalize on the promises of intellectual humility in boosting collective wisdom in social networks.

\end{abstract}

\section{Introduction}

\textit{``How many illegal immigrants are in the United States?''}---At a time when illegal immigration is at the center of national debate in the US, people's collectively held beliefs (``wisdom'') on such factual questions can substantially influence democratic processes and resulting policies. Beliefs do not form in isolation in human societies: people observe the beliefs of other group members, exchange information or evidence, and revise their own beliefs when convinced~\cite{jayles2017social,proma2024exploring}. In such complex settings, people's belief dynamics can be shaped by cognitive biases (e.g., people resist cognitive dissonance), social biases (e.g., people selectively accept information from like-minded peers), and other factors~\cite{scheffer2022belief,gawronski2012back}. These biases often result in people overestimating and underestimating the true values~\cite{gounaridis2024social}. Moreover, in the case of politically charged facts, partisan biases can shape people's beliefs and fuel political polarization~\cite{taber2006motivated,bullock2013partisan,bartels2002beyond}. Reducing the errors and polarization of collectively shared beliefs remain critical research goals.

Amid these challenges, intellectual humility (IH) emerges as a promising solution with the potential to counteract polarization and enhance the collective wisdom of humans. IH is defined by an awareness of one’s knowledge limitations, a willingness to revise one’s beliefs when warranted, and valuing other people's opinions \cite{church2016doxastic, hopkin2014intellectual, leary2017cognitive, porter2018intellectual, whitcomb2017intellectual, porter2022predictors, leary2018psychology}. In the present work, we draw from the extant literature on intellectual humility and collective intelligence and investigate \textit{whether} and \textit{how} IH can modulate the collective wisdom of humans in social networks.

\subsection{Social Influence and the Collective Wisdom of Networks}

Research on the ``Wisdom of Crowds'' has long established that the aggregated beliefs of large groups can be factually more accurate (``wise'') than the beliefs of individual group members~\cite{galton1907vox,pennycook2019fighting,wolf2015collective,surowiecki2004wisdom}. The statistical intuition is that \textit{independent} estimates from a large number of unconnected individuals can have uncorrelated or negatively correlated errors, which cancel each other out and allow the group’s collective estimate to approach the actual value~\cite{nash2014curious,sunstein2006infotopia,page2008difference}.

When individuals \textit{interact}, however, the independence property no longer holds, and people's errors become correlated. As the \textit{DeGroot process} explains, people typically revise their beliefs as a weighted average of their own beliefs and the beliefs of their social connections~\cite{degroot1974reaching}. Nevertheless, this loss of independence among estimates does not necessarily imply that collective wisdom is worse in the presence of social interactions; in fact, social influence has been shown to both elevate~\cite{becker2019wisdom,becker2017network,almaatouq2020adaptive} and harm~\cite{muchnik2013social,hahn2019communication} collective accuracies in networks based on the context. The \textit{Diversity Prediction Theorem} gives us a general rule: collective accuracy will improve if the benefits of social interaction (to individual accuracy) outweigh its drawbacks (of inducing non-independence among estimates). The theorem shows that collective error squared equals the difference between the average individual error squared and the diversity of individual judgments~\cite{page2008difference}. Thus, social influence can be advantageous if it reduces average individual error and disadvantageous if it reduces estimation diversity.

Researchers have documented various network-level and individual-level processes that modulate collective accuracies in real social networks. At a \textit{network} level, the structure of social interactions matters: for instance, decentralized networks can elevate collective wisdom over centralized ones~\cite{becker2017network}. Moreover, network structures can show temporal plasticity to amplify the impacts of high-performing individuals~\cite{almaatouq2020adaptive}. At an \textit{individual node} level, more accurate individuals tend to revise their answers less in response to social information (i.e., they place more weight on their own beliefs in the DeGroot process)~\cite{becker2017network, madirolas2015improving}. This mechanism is captured by the \textit{Revision Coefficient}, which typically shows a positive partial correlation between people's initial errors and the adjustments they make to their initial estimates, after accounting for social influence effects. This property helps pull the mean collective belief closer to the actual answer. 

Intuitively, in addition to estimation accuracy, people's psycho-social attributes can also modulate how much weight they assign to their own (and peers') estimates. In politically charged contexts, people's extraordinary preference for cognitive consistency means they disproportionately accept information from `in-group' members who already share similar beliefs~\cite{taber2006motivated,bullock2013partisan}. Such partisan biases can persist even when people are offered financial incentives for accurate estimates~\cite{guilbeault2018social}. In this work, we are interested in how the understudied yet highly promising psychological construct of \textit{intellectual humility} modulates the weights people assign to their own (and peers') estimates and how that process, in turn, affects the collective wisdom of social networks.

\subsection{Intellectual Humility}
Intellectual humility (IH) has been defined in several ways. Most definitions converge on the notion that IH involves an awareness of one's knowledge limitations and an openness to revising one's views when presented with strong evidence~\cite{church2016doxastic,hopkin2014intellectual,leary2017cognitive,porter2018intellectual,whitcomb2017intellectual}. At an \textit{individual} level, IH involves attentiveness to the limitations of (i)~the evidence supporting one's beliefs and also (ii)~one's ability to evaluate that evidence accurately. This awareness distinguishes IH from traits like indecisiveness or low self-confidence as it involves holding one's beliefs with the understanding that new evidence could reveal flaws in those beliefs~\cite{leary2018psychology}.

In addition to its individual-level cognitive features, IH has significant \textit{social} or \textit{interpersonal} qualities. Research suggests that IH is linked to greater empathy, tolerance, and positive interpersonal relationships~\cite{krumrei2017intellectual,porter2018intellectual}. People with high IH spend more time thinking about beliefs with which others disagree~\cite{krumrei2016development, leary2017cognitive, porter2018intellectual}. 
By recognizing that their beliefs might be wrong, individuals with high IH are less likely to view disagreements as personal threats and less likely to react defensively when their beliefs are challenged \cite{leary2017cognitive}. Instead, they are more likely to be open to alternative viewpoints, appreciate the intellectual strengths of others, engage constructively with differing perspectives, and reconsider their positions when presented with contrary evidence~\cite{porter2018intellectual,hopkin2014intellectual}. This relational aspect of IH is critical in environments where collaboration and consensus are essential, such as in political wisdom settings and organizational decision-making.

\subsection{The Present Work: Impact of Intellectual Humility on Collective Wisdom in Networks}

Despite the intuitively promising link between collective wisdom and intellectual humility, the two research bodies have thus far evolved fairly separately, with no notable handshake. The IH research body has overwhelmingly focused on individual-level effects, leaving a gap in our understanding of its effects on networked interaction settings~\cite{krumrei2024toward}. The collective intelligence research, in contrast, has not explicitly probed how IH can modulate the complex dynamics of belief evolution in social networks.

In this paper, we use agent-based modeling and data-calibrated simulations to explore the effects of IH on networked settings, specifically within political intelligence scenarios. We report a series of insights:

\begin{itemize}
    \item The errors and polarization of people's estimations significantly reduce in IH-enhanced treatment networks compared to non-enhanced control networks;
    \item This performance advantage can be partly explained by the significant increase in Revision Coefficients in IH-enhanced treatment networks;
    \item Both the average individual error squared and the individual estimation diversity reduce in IH-enhanced treatment networks. However, the latter reduces less, giving an overall performance advantage in terms of collective error squared based on the Diversity Prediction Theorem;
    \item The insights are robust to different network structures and task contexts.
\end{itemize}

\section{Theoretical Modeling and Simulation}

We build on DeGroot's formalization of local information aggregation, in which networked agents estimate some unknown number~\cite{degroot1974reaching}. In the classic formulation, agent $i$ updates their estimate after being exposed to the estimations of their network neighbors, $j\in \mathcal{N}_i$. In doing so, each agent adopts a weighted mean of their own estimate and the average estimate of their neighbors based on the rule:
\begin{equation}
x_{i,t+1} = \alpha_i x_{i,t} + (1 - \alpha_i)\bar{x}_{j \in \mathcal{N}_i,t}
\label{degroot}
\end{equation}
where $x_{i,t}$ is agent $i$'s estimate at time $t$; $\bar{x}_{j \in \mathcal{N}_i,t}$ is the average estimate of agent $i$'s neighbors at time $t$; $\alpha_i$ and its complement $(1 - \alpha_i)$ respectively denote the weights agent $i$ places on its own estimate and those of its peers; and $x_{i,t+1}$ denotes agent $i$'s updated estimate at time $t+1$. 

\begin{table*}[t]
\begin{center} 
\caption{Datasets used for calibrating our simulations. $N_{q}$ = total number of questions, $N_u$ = total number of unique users, $N$ = total number of data points.} 
\label{dataset} 
\begin{tabular}{lccccccc}
\hline
\textbf{Dataset} & \textbf{$N_{q}$} & \textbf{$N_{u}$} & \textbf{$N$} & \textbf{True val.}   & \textbf{Initial estimate}  & \textbf{Alpha} & \textbf{Party}  \\ 
\hline
\citeA{becker2019wisdom} (Experiment 2, social condition) & 4 & 609 & 1799  & \checkmark& \checkmark & \checkmark    & \checkmark \\
\citeA{becker2017network} (All social network conditions) & 34 & 992 & 3603  & \checkmark& \checkmark& \checkmark & \ding{55}     \\
\citeA{gurccay2015power} & 16 & 92 & 1423 & \checkmark & \checkmark & \ding{55}   & \ding{55} \\
\citeA{lorenz2011social}& 6 & 144 & 864  & \checkmark& \checkmark& \ding{55}    & \ding{55} \\ 
\hline
\end{tabular} 
\end{center} 
\end{table*}

In this model, the collective-level outcomes are determined by three intertwined factors: (i)~the distribution of initial estimates, $x_{i,t=1}$, (ii)~the distribution of self-weights, $\alpha_i$, and (iii)~the network structure of who can observe whom~\cite{becker2017network}. It has been shown that different populations (with different distributions of initial estimations and self-weights) faced with different questions and network structures can perform differently~\cite{almaatouq2022distribution}. We calibrate our simulations to the data from four previously published studies to forego assumptions about task contexts, initial estimate distributions, and self-weight distributions~\cite{becker2019wisdom, becker2017network, gurccay2015power, lorenz2011social}. These datasets contain estimations from $N_u=1837$ individuals for $N_q=60$ different questions (along with the ground truth answers), and some of these datasets contain the individuals' self-weights and party affiliation information (Table~\ref{dataset}). We experiment with $\mathcal{G} = \{$egalitarian, power-law, small-world, star$\}$ networks to confirm robustness across network structures~\cite{steger1999generating,barabasi1999emergence,watts1998collective,becker2017network}.

To initialize our model, we generate network instances with one of the structures from $\mathcal{G}$ and endow each agent, $i$, an initial estimate, self-weight, and party affiliation label sampled from a single user for a single question, $q$, in our datasets. We also assign the agent an `evidence quality' score (described below). We then create matched \textit{control} and \textit{treatment} networks, where the agents in the treatment networks enjoy enhanced intellectual humility qualities compared to their matched counterparts in the non-enhanced control networks. The agents' intellectual humility qualities are encoded in their modulated self-weights, $\tilde{\alpha}_i^{(c)}$, where $c \in \mathcal{C}$ denotes a condition in $\mathcal{C}=\{$control, treatment$\}$. The agents then communicate with their network peers across $t=1,2,3$, where $t=1$ indicates the initial condition with identical estimates for control and treatment agents. Each agent then revises their estimate twice at $t=2,3$ using the DeGroot rule in Equation~\ref{degroot}. This matched setup allows us to compare the benefits of intellectual humility between the two conditions. We explain how IH qualities are captured in self-weights below.

\subsection{Capturing IH in Self-weight Parameters}
The self-weight parameter, $\alpha_i$, governs the importance agent $i$ gives to its own estimates relative to the peers' estimates. We take the self-weight sampled from the calibration datasets, $\alpha^0_i$, as the baseline self-weight for the agent $i$. It is well-known that more accurate individuals revise their responses less (i.e., have higher self-weights)~\cite{burton2021rewiring}. This process is corroborated in all our datasets: for example, the Pearson's correlation between $\alpha^0_i$ and the initial error of agent $i$ ranges between $-0.12$ and $-0.19$ for the four questions in \citeA{becker2019wisdom} ($P<0.001$ in all cases). Thus, $\alpha^0_i$ contains some information about the agent's understanding of their own accuracy. In creating the matched networks for $c \in \mathcal{C}$, we modulate the baseline $\alpha^0_i$ as,
\begin{equation}
    \tilde{\alpha}_i^{(c)} = \alpha^0_i + \triangle \alpha_i^{(c)}
    \label{alpha_update_eq}
\end{equation}
where $\triangle \alpha_i^{(c)}$ captures the changes to $\alpha^0_i$ based on condition $c$. To capture this change, we borrow two properties from IH theory regarding how intellectually humble treatment agents respond differently to their peers' \textit{evidence quality} and \textit{homophily} signals compared to their control counterparts.

In real-world discussions, people, more often than not, exchange evidence, contextual information, or reasoning relevant to the wisdom scenario (referred to with an `evidence quality' score). If an agent \textit{perceives} the peers' aggregated evidence quality signals to be convincing, we can expect the agent to pay more attention to the peers (i.e., reduce self-weight) and vice versa. However, the \textit{perceived} evidence quality is not the same as the \textit{objective} evidence quality---since the receiver's cognitive biases (e.g., confirmation bias), social biases (e.g., partisan homophily), and other factors can distort their perception of the incoming objective evidence quality~\cite{scheffer2022belief}. Crucially, IH can affect how well an agent can (i)~evaluate a piece of evidence and (ii)~resist partisan homophily biases---which, in turn, can affect their modulations of baseline self-weights. This suggests that $ \triangle \alpha_i^{(c)}$ can be related to perceived evidence qualities. We decompose $ \triangle \alpha_i^{(c)}$ into a linear combination of objective evidence quality and homophily-bias-based effects.

\subsubsection{Effects of Evidence Quality Evaluation.} Individuals with higher IH are more likely to evaluate the strength of evidence critically, paying close attention to the evidential basis of their own pre-existing beliefs and the socially received information~\cite{leary2018psychology, krumrei2020intellectual, zmigrod2019psychological}. This reflective process helps individuals balance open-mindedness and discernment to prioritize valuable information selectively~\cite{vorobej2011distant,leary2022intellectual,koriat1980reasons,kuhn1994well,kahan2013ideology,kunda1990case}. Thus, treatment agents with high IH can be endowed with the ability to evaluate the quality of evidence better---both their own and that of others---compared to their matched counterpart agents in the control network. This effect can transpire even if the agents have no information about their peers' political alignments.

The users in our calibration datasets never gave any evidence supporting their estimates. Therefore, in our simulations, we assign each agent a synthetic objective evidence quality score, $d$, correlating with their initial estimation accuracy with added Gaussian noise. This stochastic linear relation is based on the observation that highly accurate users tend to show some understanding of their own accuracy, potentially allowing them to provide better evidentiary support when needed. If the peers' average evidence quality surpasses the agent's own, an intellectually humble agent will reduce their self-weight more than the control, reflecting their openness to learning from more potent viewpoints. Conversely, if the peers' average evidence quality is lower than the agent's own, an intellectually humble agent will increase their self-weight more than the control, signifying reliance on their own perspective. This mechanism models IH as a dynamic process driven by evidence-based discernment rather than indiscriminate openness. Thus, an agent's self-weight modulation can be partly related to the difference between their own evidence quality, $d_i$, and the peers' average evidence qualities, $\bar{d}_{j \in \mathcal{N}_i}$:
\begin{equation}
    \triangle \alpha_{i}^{'(c)} = w_{1}^{(c)} + w_{2}^{(c)}(d_{i} - \bar{d}_{j \in \mathcal{N}_i})
    \label{evidence_eq}
\end{equation}
where, $w_{1}^{(c)}$ is the intercept and $w_{2}^{(c)}$ is the \textit{positive} slope of the linear relation. $w_2^{(c)}$ is steeper for treatment networks than control networks to capture the superior evidence evaluation capabilities of the treatment agents.

\subsubsection{Effects of Homophily.} Homophily refers to the tendency of individuals to associate with others who share similar beliefs, values, or affiliations \cite{homer2013complex,mcpherson2001birds,scheffer2003slow}. In the case of political wisdom, being aware of a peer's political alignment can lead to partisan homophily-based information prioritization. This can induce correlated errors and polarization in people's estimates~\cite{becker2019wisdom,esteve2022homophily,golub2012homophily}. However, high-IH individuals are more likely to value dissenting viewpoints, engage constructively and tolerantly with out-group perspectives, and reconsider their own positions when warranted~\cite{leary2018psychology,porter2022predictors,sgambati2022intellectual,krumrei2020intellectual, krumrei2021sociopolitical}. Thus, in the presence of party alignment information, treatment agents with high IH can be simulated to resist the homophily-based social reinforcements better and be more receptive to differing opinions than their control counterparts. Namely, agent $i$'s self-weight modulation can be partly related to the fraction of its peers who share the same political alignment, $h_i$, 
\begin{equation}
    \triangle \alpha_{i}^{''(c)} = w_{3}^{(c)} + w_{4}^{(c)}h_i
    \label{hom_eq}
\end{equation}
where, $w_{3}^{(c)}$ is the intercept and $w_{4}^{(c)}$ is the \textit{negative} slope of the linear relation. $w_4^{(c)}$ is less steep for treatment networks than control networks to capture the superior ability of the treatment agents to ignore partisan bias.

We model the self-weight modulation, $\triangle \alpha_i^{(c)}$, as a linear combination of $\triangle \alpha_{i}^{'(c)}$ and $\triangle \alpha_{i}^{''(c)}$. From Equations~\ref{alpha_update_eq}, \ref{evidence_eq}, and \ref{hom_eq}, we get the expression for modulated self-weight, $\tilde{\alpha}_i^{(c)}$,
\begin{equation}
    \tilde{\alpha}_i^{(c)} = \alpha^0_i + w_{1}^{(c)} + w_{2}^{(c)}(d_{i} - \bar{d}_{j \in \mathcal{N}_i}) + w_{3}^{(c)} + w_{4}^{(c)}h_i
    \label{alpha_update_eq_final}
\end{equation}
We clip $\tilde{\alpha}_i^{(c)}$ to bind it between $0$ and $1$, a property observed to hold for nearly all users in our dataset.

\subsection{Metrics}
\subsubsection{Individual Error.} To facilitate comparisons across different estimation tasks of different scales, we report errors in terms of the number of standard deviations an estimate is away from the true answer (comparable to a $z$-score)~\cite{becker2017network}. Namely, the `individual error' is calculated as the absolute distance between the estimate and the truth, normalized by the standard deviation: 
\begin{equation}
e_{i,q,t} =  \frac{|x_{i,q,t} - x^*_{q}|} {\sigma_{q,t=1}}
\label{error_eq}
\end{equation}
where, $x_{i,q,t}$ and $e_{i,q,t}$ respectively denote agent $i$'s estimate and error in response to question $q$ at time $t$; $x^*_q$ is the true value of question $q$; and $\sigma_{q,t=1}$ is the standard deviation of the baseline responses to question $q$ at time $t=1$.

\subsubsection{Polarization.} We first calculate the absolute distance between the SD-normalized estimates of every possible cross-party pair of Republicans and Democrats for a single question. We then calculate the mean of these values to get the average pairwise distance (`polarization') for the question:
\begin{equation}
p_{q,t} = \frac{1}{n^{(D)} n^{(R)}} \sum_{i=1}^{n^{(D)}} \sum_{j=1}^{n^{(R)}} \frac{| x_{i,q,t}^{(D)} - x_{j,q,t}^{(R)}|} {\sigma_{q,t=1}}
\label{polarization_eq}
\end{equation}
where, $p_{q,t}$ is the polarization (average pairwise distance) for question $q$ at time $t$; $n^{(D)}$ and $n^{(R)}$ respectively denote the number of agents from the Democratic ($D$) and Republican ($R$) parties; and $x_{i,q,t}^{(k)}$ is agent $i$'s estimate for question $q$ at time $t$, where the agent belongs to the political party, $k\in \{D,R\}$.

\begin{figure*}[t]
    \centering
    \includegraphics[width=1\linewidth]{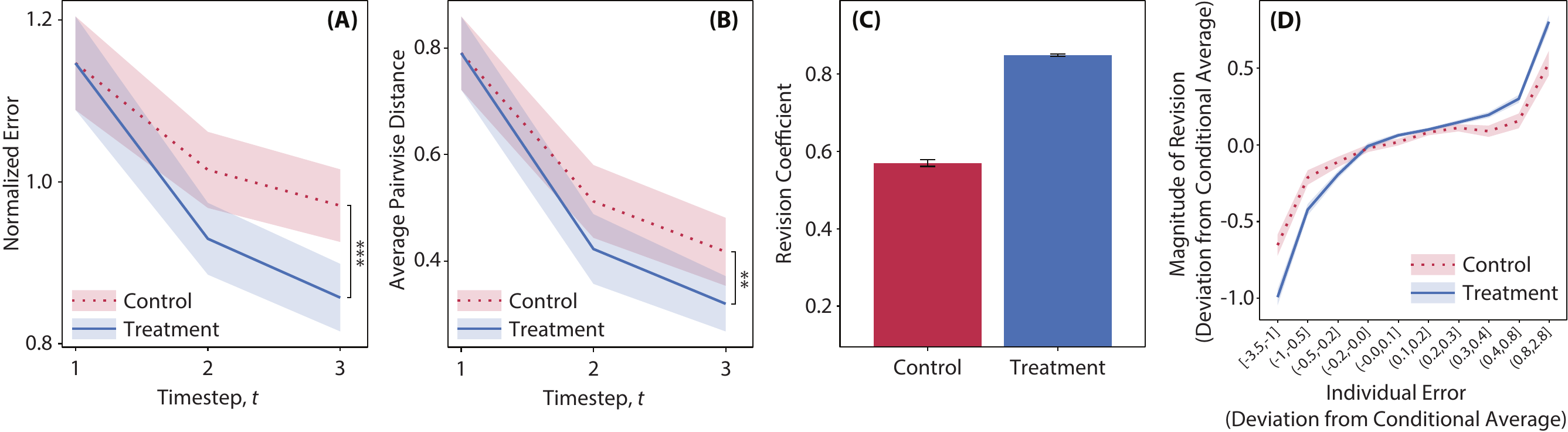}
    \caption{\textbf{(A)} Normalized estimation errors and \textbf{(B)} Polarization (average pairwise distance) significantly reduce in treatment networks, while \textbf{(C)} Revision Coefficients increase. \textbf{(D)} The partial correlation between  $\tilde{e}_{i,q}$ and $\triangle \tilde{x}_{i,q}$ (adjusted for $s_{i,q}$) improves in the treatment condition. Shaded regions and whiskers denote $95\%$ C.I. *** $P<0.001$, ** $P<0.01$.}
    \label{pnas_2019_fig}
\end{figure*}

\subsubsection{Revision Coefficient.} We calculate the Revision Coefficient as the partial correlation between the \textit{individual initial errors} ($e_{i,q,t=1} = | x_{i,q,t=1} - x_q^*|/\sigma_{q,t=1}$) and the \textit{individual revisions} ($\triangle x_{i,q} = |x_{i,q,t=3} - x_{i,q,t=1}|/\sigma_{q,t=1}$) of the agents, adjusted for the \textit{social signals} ($s_{i,q} = | x_{i,q,t=1} - \bar{x}_{j \in \mathcal{N}_i,q,t=1}|/\sigma_{q,t=1}$) that the agents receive. Adjusting for the social signal is important as individuals with higher initial errors often receive stronger social feedback as they deviate from the rest. The `Revision Coefficient' is calculated as, 
\begin{equation}
r_{q} = corr(\tilde{e}_{i,q}, \triangle \tilde{x}_{i,q})
\label{rev_coeff_eq}
\end{equation}
where, $r_q$ denotes the Revision Coefficient for question $q$; $\tilde{e}_{i,q}$ denotes individual initial error ($e_{i,q,t=1}$) adjusted for social signal ($s_{i,q}$); and $\triangle \tilde{x}_{i,q}$ denotes the individual revision ($\triangle x_{i,q}$) adjusted for social signal ($s_{i,q}$).

\subsubsection{Notations for the Diversity Prediction Theorem.} The Diversity Prediction Theorem~\cite{page2008difference} shows that the collective error squared ($SqE^{(collective)}$), the average individual error squared ($SqE^{(individual)}$), and the predictive diversity of individual judgments ($PD$) are related as,
\begin{equation}
SqE^{(collective)} = SqE^{(individual)} - PD
\label{div_theorem}
\end{equation}
where,
\begin{equation}
SqE^{(collective)} = (\bar{x} - x^* )^2
\label{div_theorem_sqec}
\end{equation}
\begin{equation}
SqE^{(individual)} = \frac{1}{n_u} \sum_{i=1}^{n_u} (x_i - x^* )^2
\label{div_theorem_sqei}
\end{equation}
\begin{equation}
PD =  \frac{1}{n_u} \sum_{i=1}^{n_u} (x_i - \bar{x} )^2
\label{div_theorem_pd}
\end{equation}
with $n_u$ denoting the number of agents in a given network (from a single question) and $\bar{x}$ denoting the mean estimate of all those agents.

\begin{table*}[t]
\begin{center} 
\caption{Robustness tests against different datasets and network structures. The main result setting is italicized. For brevity and clarity, we use \checkmark to denote results with the same directions as the main results. Significance codes: *** $P<0.001$, ** $P<0.01$.} \label{robustness_table}
\begin{tabular}{lccccc}
\hline

\textbf{Dataset} & \textbf{Network Structure} &  \textbf{Error} & \textbf{Rev Coeff} & \textbf{Polarization} \\
\hline
\multirow{2}{*}{\citeA{becker2019wisdom}}  & \textit{Egalitarian }             & \textit{\checkmark}***  & \textit{\checkmark}*** & \textit{\checkmark}** \\
     & Barabási–Albert               & \checkmark***  & \checkmark*** & \checkmark*** \\ 
      & Watts--Strogatz              & \checkmark***  & \checkmark*** & \checkmark*** \\
      & Star               & \checkmark***  & \checkmark*** & \checkmark*** \\ \hline
\multirow{2}{*}{\citeA{becker2017network}} 
    &   Egalitarian         & \checkmark***  & \checkmark*** & - \\
     & Barabási–Albert         & \checkmark***  & \checkmark*** & - \\ 
     & Watts--Strogatz        & \checkmark***  & \checkmark*** & - \\ 
     & Star         & \checkmark***  & \checkmark*** & - \\ \hline
\multirow{2}{*}{\citeA{gurccay2015power}}       
    &   Egalitarian        & \checkmark***  & \checkmark*** & - \\
     & Barabási–Albert       & \checkmark***  & \checkmark*** & - \\
     & Watts--Strogatz       & \checkmark***  & \checkmark*** & - \\
     & Star        & \checkmark***  & \checkmark*** & - \\ \hline
\multirow{2}{*}{\citeA{lorenz2011social}}  
    &   Egalitarian         & \checkmark***  & \checkmark*** & -\\
     & Barabási–Albert          & \checkmark***  & \checkmark*** & - \\
     & Watts--Strogatz        & \checkmark***  & \checkmark*** & - \\
     & Star         & \checkmark**  & \checkmark*** & - \\
\hline
\end{tabular}
\end{center}
\end{table*}

\section{Results}
We report the main results for the~\citeA{becker2019wisdom} dataset since it contains politically charged questions and the users' party affiliation information (Table~\ref{dataset}). We conduct $100$ replications for each network structure in $\mathcal{G}$, calibrated to the real users' data from a single question $q$. We assign the users different node positions randomly and independently in each replication and aggregate each agent's estimate updates across all replications (we use the egalitarian network structure for the main results and test robustness for the other structures). We use the remaining three datasets for robustness tests~\cite{becker2017network,gurccay2015power,lorenz2011social}. Since these three datasets do not have party affiliation information, we set the homophily component $\triangle \alpha_{i}^{''(c)}=0$ in Equation~\ref{alpha_update_eq_final}. When the baseline self-weights are absent in a dataset, we follow~\citeA{burton2021rewiring} to synthesize $\alpha_i^{0}$ values using a stochastic linear relationship with agent $i$'s initial estimation errors.

\subsection{Estimation Errors and Polarization Reduce in Intellectually Humble Networks}
We find that the treatment networks consistently exhibit significantly lower errors than the control networks across all estimation tasks ($\beta=-0.11, SE = 0.01, P<0.001$, linear mixed effects model with user ID, question ID, and party ID captured as random effects; Figure~\ref{pnas_2019_fig}(A)). We use a linear mixed-effects model to account for all repeated measures and dependencies among data points. 

We similarly analyze the effect of IH on polarization. The treatment networks show significantly lower polarization than the control networks ($\beta=-0.09, SE = 0.02, P<0.01$, linear mixed effects model with question ID captured as random effects; Figure~\ref{pnas_2019_fig}(B)).

We examine the collective outcomes using $SqE^{(collective)}$. We find treatment $SqE^{(collective)}$ to be $12.6\%$ lower than the control. Using the Diversity Prediction Theorem to decompose $SqE^{(collective)}$, we observe that treatment $SqE^{(individual)}$ is $19.5\%$ lower than the control. However, this performance gain is slightly counteracted by a loss of diversity, as treatment $PD$ also reduces by $6.9\%$. Since $PD$ reduces less, we get an overall performance advantage at the collective level.

\subsection{The Moderating Role of Revision Coefficients}
The treatment networks show significantly higher Revision Coefficients than the control networks ($\beta=0.41, SE = 0.004, P<0.001$, linear mixed effects model with question ID and party ID as random effects; Figure~\ref{pnas_2019_fig}(C)). This effect is further illustrated in Figure~\ref{pnas_2019_fig}(D), which shows the partial correlation between $\tilde{e}_{i,q}$ and $\triangle \tilde{x}_{i,q}$ (adjusted for $s_{i,q}$). As expected, the control agents revise their estimates less when they are more accurate, after accounting for social influence. The treatment agents, however, can better discern evidence qualities and resist partisan homophily biases. This results in a steeper slope of the treatment curve in Figure~\ref{pnas_2019_fig}(D), increasing the Revision Coefficient and pulling the group estimations closer to the truth. This insight suggests a mechanism behind how IH can moderate wisdom in social networks.

\subsection{Robustness}

The $N_q=60$ questions in our four datasets have true values ranging between $-46$ and $7,825,200$, and our results are robust to all questions (Table~\ref{robustness_table}). The results are also robust to different network structure choices: Egalitarian ($d=4, n=n_u$)~\cite{steger1999generating}, Barabási–Albert's power law ($n=n_u, m=2, p=0.5$)~\cite{barabasi1999emergence}, Watts-Strogatz's small-world ($n=n_u, k=4, p=0.5$)~\cite{watts1998collective}, and centralized star network structures ($n=n_u$)~\cite{becker2017network}. Furthermore, we test robustness against the fraction of agents in the treatment condition who can successfully demonstrate IH in their estimation revisions. In this case, a random $f\%$ of the treatment agents are endowed with enhanced IH while the remaining agents remain identical to their control counterparts. Simulations show that the main results remain consistent even when $f$ is as low as $10\%$, with statistical significance emerging for $f$ between $20\%$ and $60\%$ for different tasks and network contexts.

\section{Discussion}

This work contributes to the body of collective wisdom research, illustrating \textit{whether} and \textit{how} the psychological construct of intellectual humility can affect the errors and polarization of collectively held beliefs. Our model encodes IH in the self-weight parameters of the DeGroot process. Using this model, we show that IH-enhanced networks exhibit lower errors and polarization, with robust results across tasks and network structures. Our simulations capture how IH fosters balanced belief revisions: less accurate individuals revise more effectively, while more accurate individuals better retain their performance. This effect is showcased by an elevated Revision Coefficient in the IH-enhanced networks, suggesting a mechanism driving the collective performance benefits.

A body of agent-based simulations previously probed how psycho-social elements (e.g., mutual trust) and network structural elements (e.g., network size and connectivity) can modulate polarization, echo chambers, and inaccuracies in epistemic interaction-based social networks~\cite{olsson2013bayesian,madsen2018large,hahn2020truth,franken2021}. For instance, when Bayesian agents become trusting of their peers' credibility, a snowball effect can ensue that polarizes even rational agents~\cite{olsson2013bayesian,franken2021}. Moreover, network structural properties can aid echo chamber formation, while cognitive and social biases can exacerbate the process~\cite{madsen2018large,hahn2020truth}. Our work augments this research thread by showing that IH can reduce polarization and collective errors by encouraging \textit{balanced} belief revisions, and that the effects remain robust across network structures.

Our results also corroborate and extend existing IH research that hypothesizes higher accuracy and reduced extremism at the individual level~\cite{knochelmann2024effects,smith2023you,bowes2021looking}. We show that these benefits can manifest in closed social systems if the interactants' IH can be practically elevated. However, designing a \textit{scalable} intervention for elevating IH in a network setting is non-trivial and formulates part of our future work. Such empirical data can help validate and extend our model by elucidating further nuances of IH. This research direction holds immense potential for informing applications in organizational decision-making, public policy, and collaborative problem-solving---not to mention in politically charged contexts for combating collective inaccuracies in societies.

\section{Code Availability}
The code and data for replicating the work can be found here: https://github.com/cssai-research/wisdom-ih-networks.

\newpage
\section{Acknowledgments}
A faculty startup fund from the University of South Florida supported this work.

\setlength{\bibleftmargin}{.125in}
\setlength{\bibindent}{-\bibleftmargin}

\end{document}